\documentclass[aps,floatfix]{revtex4}
\usepackage{graphics,epsfig}
\usepackage{graphicx}

\begin{document}

\title{Maximally extended, explicit
and regular coverings of the \\Schwarzschild - de Sitter vacua in
arbitrary dimension}
\author{Kayll Lake \cite{email}}
\affiliation{Department of Physics and Department of Mathematics
and Statistics, Queen's University, Kingston, Ontario, Canada, K7L
3N6 }
\date{\today}

\begin{abstract}
Maximally extended, explicit and regular coverings of the
Schwarzschild - de Sitter family of vacua are given, first in
spacetime (generalizing a result due to Israel) and then for all
dimensions $D$ (assuming a $D-2$ sphere). It is shown that these
coordinates offer important advantages over the well known Kruskal
- Szekeres procedure.
\end{abstract}
\maketitle
\section{Introduction}
A maximally extended and regular covering of the Schwarzschild
vacuum is now a fundamental part of any introduction to general
relativity. Almost always the covering is given by way of the
implicit Kruskal \cite{kruskal} - Szekeres \cite{szekeres}
procedure \cite{procedure}. However, as emphasized long ago by
Ehlers \cite{ehlers}, an explicit maximally extended and regular
covering of the Schwarzschild vacuum is known and was first given
by Israel \cite{israel}. Unfortunately, despite the fact that
these coordinates offer many advantages over the Kruskal -
Szekeres coordinates, they are almost never used. In this paper I
extend Israel's procedure to the Schwarzschild - de Sitter class
of vacua in four dimensions and then to arbitrary dimensions $D$
assuming a $D-2$ sphere as in the Tangherlini generalization of
the Schwarzschild vacuum. Moreover, it is made clear that these
coordinates offer important advantages over the Kruskal - Szekeres
procedure. These advantages include: An explicit representation of
the line element that can be extended to arbitrary dimension, a
simultaneous covering of both the black hole and cosmological
horizons and derivation by  direct integration of Einstein's
equations \cite{transformations} without recourse to coordinate
transformations.
\section{Coordinate construction (D=4)}
We start with a spherically symmetric spacetime in coordinates
$(u,w,\theta,\phi)$ where $k^{\alpha}=\delta^{\alpha}_{w}$ is a
null four-vector so that the line element takes the form
\cite{conventions}
\begin{equation}
ds^2=f(u,w)du^2+2h(u,w)dudw+r(u,w)^2d\Omega^2\label{generalmetric}
\end{equation}
where $d\Omega^2$ is the metric of a unit two-sphere ($d
\theta^2+\sin^2\theta d\phi^2$). Further, setting
$k^{\beta}\nabla_{\beta}k^{\alpha}=0$ (so that trajectories of
constant $u, \theta$ and $\phi$ are radial null geodesics affinely
parameterized by $w$) it follows that $\partial h/\partial w=0$.
Defining $U \equiv \int h(u) du$ and rewriting $U$ as $u$ we set
$h=1$ in (\ref{generalmetric}) which remains a completely general
spherically symmetric spacetime. The expansion of $k^{\alpha}$
reduces to
\begin{equation}
\nabla_{\alpha}k^{\alpha}=\frac{2}{r}\frac{ \partial r}{\partial
w}. \label{expansiongen}
\end{equation}
Associated with the vector field $l^{\alpha}=\delta^{\alpha}_{u}$
we have $l^{\alpha}l_{\alpha}=f$,
\begin{equation}
l^{\beta}\nabla_{\beta}l^{\alpha}=\frac{1}{2}(-\frac{
\partial f}{\partial w},\frac{
\partial f}{\partial u}+f\frac{
\partial f}{\partial w},0,0),\label{lacc}
\end{equation}
and
\begin{equation}
\nabla_{\alpha}l^{\alpha}=\frac{2}{r}\frac{
\partial r}{\partial u}.\label{lexp}
\end{equation}
The trajectories of constant $w, \theta$ and $\phi$ are radial
null geodesics (affinely parameterized by $u$) only for $f=\frac{
\partial f}{\partial u}=\frac{
\partial f}{\partial w}=0$.
Note that if $\frac{\partial f}{\partial u}=0$ then
$m^{\alpha}=(-2/f,1,0,0)$ is tangent to a radial null geodesic and
clearly $w$ is again affine. The associated geodesics immediately
follow as $u=-2\int dw/f+ \delta$ where $\delta$ is a constant.
Important examples of this special case are given below.
\section{Vacua with $\Lambda$ (D=4)}
For the spacetime (\ref{generalmetric}) with $h=1$,
\begin{equation}
f = \frac{w}{3urC^2}(-2C(C-r)^2+uw(2C+r)),\label{metric}
\end{equation}
and
\begin{equation}
r=\frac{uw(3M-C)}{C^2}+C,\label{r(u,w)}
\end{equation}
where $C$ ($ \neq 0$) and $M$ ($ \geq 0$) are constants, it
follows that
\begin{equation}
R_{\alpha}^{\beta}=\Lambda \delta_{\alpha}^{\beta},\label{ricci}
\end{equation}
\begin{equation}
R=4\Lambda,\label{ricciscalar}
\end{equation}
and
\begin{equation}
C_{\alpha \beta \gamma \delta}C^{\alpha \beta \gamma
\delta}=\frac{48 M^2}{r^6},\label{weylscalar}
\end{equation}
where
\begin{equation}
\Lambda=\frac{3(C-2M)}{C^3},\label{lambda}
\end{equation}
$R_{\alpha}^{\beta}$ is the Ricci tensor, $R$ the Ricciscalar and
$C_{\alpha \beta \gamma \delta}$ the Weyl tensor. Viewing
$\Lambda$ as a constant of nature and $M$ a property of the
vacuum, it is clear from (\ref{lambda}) that the specification of
$C$ does not determine the physical situation uniquely. Since, by
(\ref{r(u,w)}), the axes are centered on $r=C$, it is important to
distinguish ranges in $C$. Writing $C=\alpha M$ with $\alpha \neq
0$ and $M \neq 0$ it follows immediately from (\ref{lambda}) that
\begin{equation}
9 \Lambda M^2=\frac{27 (\alpha - 2)}{\alpha^3}. \label{crange}
\end{equation}
It follows from (\ref{crange}) that all values of $\Lambda M^2$
are determined uniquely by $\alpha$ using the following ranges:
$-\infty < 9 \Lambda M^2 < 1$ for $0 < \alpha < 3$ and $9 \Lambda
M^2
>1$ for $-6 < \alpha < 0$. The latter range is of no interest and is not discussed here.

\bigskip

The one independent invariant derivable from the Riemmann tensor
without differentiation can be taken to be (\ref{weylscalar}) and
so with $u$ and $w$ extending over the reals,  the vacua are
maximally extended and regular for $0< r < \infty$ (and for all
finite $r$ if $M=0$). In particular, note that
\begin{equation}
g_{uu} | _{_{r=C}}=\frac{w^2}{C^2}.\label{metricrc}
\end{equation}

With (\ref{r(u,w)}) it follows that
\begin{equation}
\nabla_{\alpha}k^{\alpha}=\frac{2u(3M-C)}{r
C^2}=\frac{u}{w}\nabla_{\alpha}l^{\alpha}.\label{expansion}
\end{equation}
From (\ref{metric}) and (\ref{r(u,w)}) it follows that
trajectories of constant $w, \theta$ and $\phi$ are radial null
geodesics only for $w=0$. More generally, the radial null
geodesics that satisfy
\begin{equation}
\frac{d w}{d u}=-\frac{1}{2}f,\label{othernull}
\end{equation}
with $f$ given by (\ref{metric}) and (\ref{r(u,w)}), can be
written down in terms of elementary functions (as explained
below). Note that for these trajectories $du/dw \rightarrow 0$ as
$r \rightarrow 0$ ($C \neq 3M$) and $dw/du \rightarrow 0$ as $w
\rightarrow 0$. We distinguish the cases: $M=0$ (de Sitter),
$C=3M$ (Bertotti-Kasner), $C=2M$ (Schwarzschild), $2M<C<3M$
(Schwarzschild-de Sitter), $0<C<2M$ (Schwarzschild-anti de Sitter)
and discuss them below. For the case $C=2M$ the coordinates were
first obtained by Israel \cite{israel}. Not covered as special
cases are anti de Sitter space (see Appendix A) and the degenerate
Schwarzschild-de Sitter spacetime (see Appendix B).

\subsection{$M=0$ (de Sitter space)}

\begin{widetext}
\begin{figure}[ht]
\epsfig{file=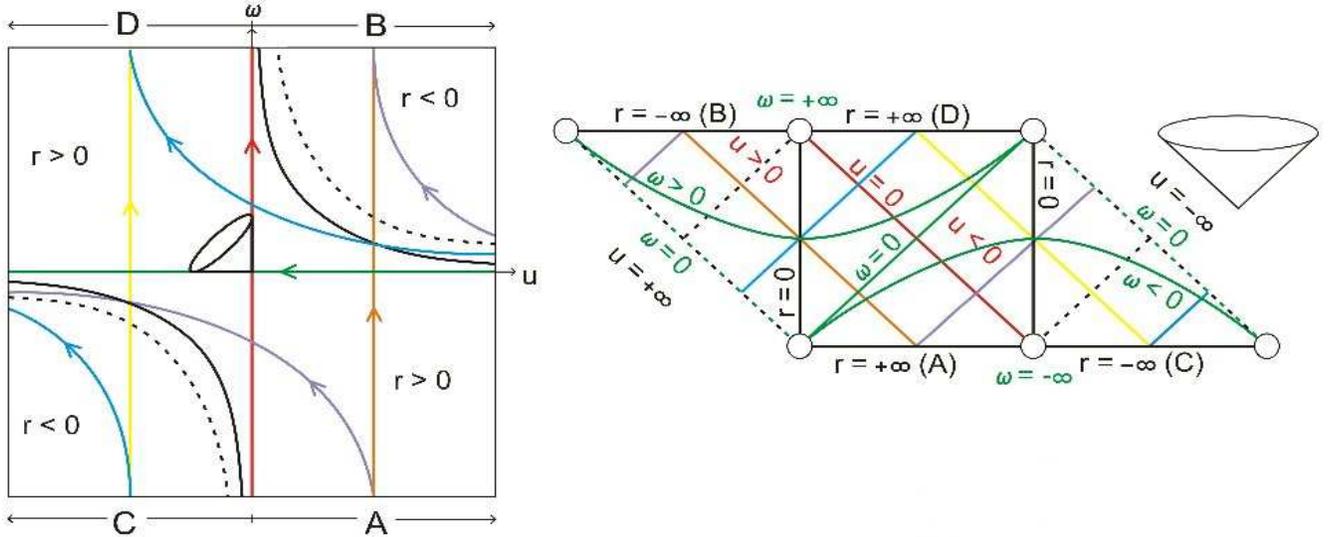,height=3in,width=7in,angle=0}
\caption{\label{desitterfig}(colour coded) At left, representation
of de Sitter space in the $u - w$ diagram. Coloured curves are
radial null geodesics all affinely parameterized by $w$. The solid
black curves give $r=0$ and the dashed black curves give
$r=-\sqrt{3/ \Lambda}$. The axes are centered on $r=\sqrt{3/
\Lambda}$. The future null cone at the origin ($u=w=0$) is shown.
The trajectories $w=$ constant are spacelike for $w\neq0$. At
right, the Penrose-Carter diagram of de Sitter space. The global
orientation of the future null cone is shown. The same scheme as
the $u - w$ diagram is used and some trajectories of constant $w$
are shown. Identification of $w=0, u=\pm\infty$ gives a null
geodesically complete causal representation of de Sitter space. }
\end{figure}
\end{widetext}

With $M=0$, $\Lambda=3/C^2$,
$r=\sqrt{\frac{\Lambda}{3}}(\frac{3}{\Lambda}-uw)$ and we have de
Sitter space. The metric simplifies to
\begin{equation}
ds^2=\frac{\Lambda}{3}w^2du^2+2dudw+\frac{\Lambda}{3}(\frac{3}{\Lambda}-uw)^2d\Omega^2.\label{desitter}
\end{equation}
Trajectories with four tangents $m^{\alpha}=(-\frac{6}{\Lambda
w^2} ,1,0,0)$ are radial null geodesics (so $w$ is affine for both
radial null directions). We can write these geodesics in the form
\begin{equation}
w=\frac{6}{\Lambda u+\delta}\label{desittergeo}
\end{equation}
where $\delta$ is a constant. The negative cosmological horizons
($r=-\sqrt{\frac{3}{\Lambda}}$) are then given by $\delta=0$ where
the expansion $\nabla_{\alpha}m^{\alpha}$ ($=\frac{2(6-uw \Lambda
)}{w(3-uw \Lambda)}$) vanishes. Some details are shown in FIG.
\ref{desitterfig}.

\subsection{$C=3M$ (Bertotti-Kasner space)}

With $C=3M=1/\sqrt{\Lambda}=r$ we have Bertotti - Kasner space.
The metric simplifies to
\begin{equation}
ds^2=\Lambda w^2
du^2+2dudw+\frac{1}{\Lambda}d\Omega^2.\label{bertottik}
\end{equation}
Trajectories with four tangents $m^{\alpha}=(-\frac{2}{\Lambda
w^2},1,0,0)$ are radial null geodesics (so again $w$ is affine for
both radial null directions) but now
$\nabla_{\alpha}k^{\alpha}=\nabla_{\alpha}m^{\alpha}=0$. We can
write these geodesics in the form
\begin{equation}
w=\frac{2}{\Lambda u+\delta}\label{bertottigeo}
\end{equation}
where $\delta$ is a constant. The $u-w$ plane is like that of de
Sitter space now with $uw=2/\Lambda$ (again $\delta=0$)
distinguishing the branches of the $m$ geodesics. This space shows
that the Birkhoff theorem does not extend directly to $\Lambda >
0$ (see the case $2M < C < 3M$ below) \cite{rindler}.

\subsection{$C=2M$ (Schwarzschild vacuum)}

\begin{widetext}
\begin{figure}[ht]
\epsfig{file=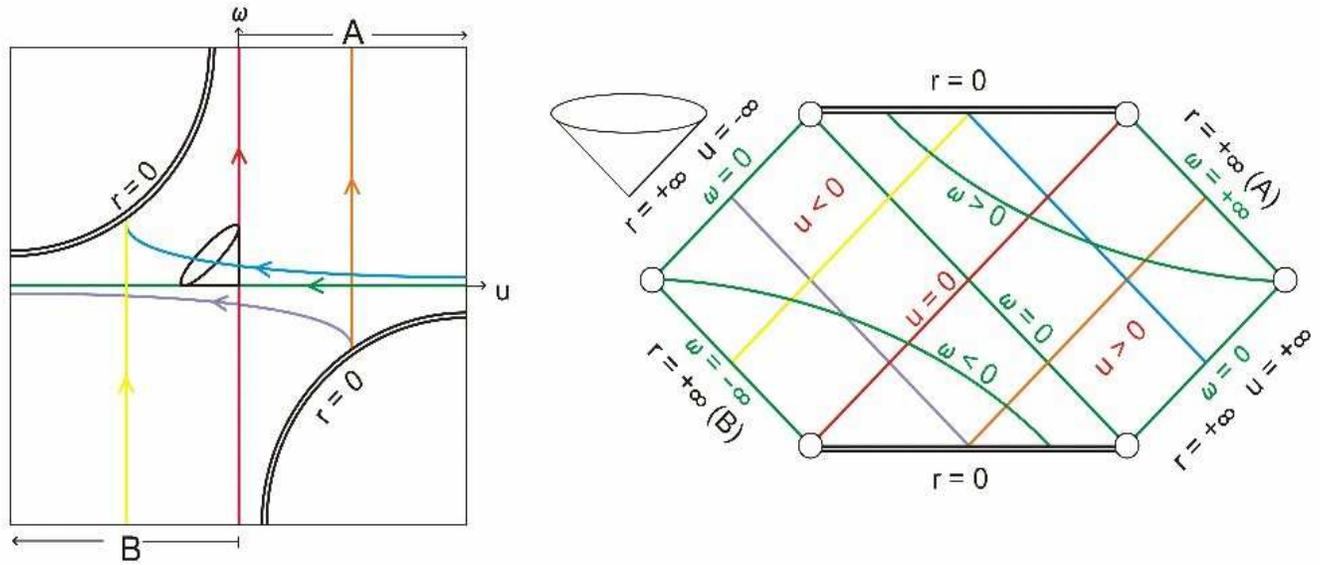,height=3in,width=7in,angle=0}
\caption{\label{schwarzschild}(colour coded) At left,
representation of the Schwarzschild vacuum in the $u - w$ diagram.
Coloured curves are radial null geodesics. The double black curves
give $r=0$. The axes are centered on $r=2M$. The future null cone
at the origin ($u=w=0$) is shown. The trajectories $w=$ constant
are spacelike for finite $w\neq0$. At right, the Penrose-Carter
diagram of the Schwarzschild vacuum. The global orientation of the
future null cone is shown. The same scheme as the $u - w$ diagram
is used and some trajectories of constant $w$ are shown. Unlike
the cases with $\Lambda \neq 0$, we do not identify $w=0,
u=\pm\infty$. The reason for this is that here $\ln w$ is an
affine parameter for the $m$ geodesics and so the diagrams are
null geodesically complete.}
\end{figure}
\end{widetext}

With $C=2M$, $\Lambda=0$, $r=\frac{8M^2+uw}{4M}$ and we have the
Schwarzschild vacuum in Israel coordinates \cite{israel}. The
metric simplifies to
\begin{equation}
ds^2=\frac{2w^2}{uw+8M^2}du^2+2dudw+(\frac{8M^2+uw}{4M})^2d\Omega^2.\label{schw}
\end{equation}
Trajectories with four tangents
$m^{\alpha}=(-\frac{uw+8M^2}{w},w,0,0)$ are radial null geodesics
so $w$ is not affine. Now
$\nabla_{\alpha}m^{\alpha}=-\frac{16M^2}{uw+8M^2}=-\frac{4M}{r}$.
We write these geodesics in the form
\begin{equation}
uw=-8M^2\ln(\delta w)\label{schwgeo}
\end{equation}
where $\delta$ is a constant. Some details are shown in FIG.
\ref{schwarzschild}.

\subsection{$2M < C < 3M$ (Schwarzschild - de Sitter)}

Again writing $C=\alpha M$ it follows that for $2 < \alpha < 3$
there is another constant $E > C$ for which
$\Lambda=\frac{3(E-2M)}{E^3}$. The constant is given by
\begin{equation}
E=\beta M \label{coshorizon}
\end{equation}
with
\begin{equation}
\beta={\frac {( 2-\alpha+\sqrt {( \alpha+6)( \alpha -2) })
\alpha}{2(\alpha-2)}}.\label{beta}
\end{equation}
The trajectories $r=E$ are radial null geodesics (with non-zero
expansion) tangent to the ``cosmological" horizons. Note that
$\beta \rightarrow 3$ as $\alpha \rightarrow 3$ and $\beta
\rightarrow\infty$ as $\alpha \rightarrow 2$. Now define
\begin{equation}
\mathcal{A}\equiv{\alpha}^{4}{M}^{2}+uw ( \alpha-2 )  ( 3-\alpha
)\label{adef}
\end{equation}
and
\begin{equation}
\mathcal{B}\equiv{\alpha}^{3}{M}^{2}+( 3- \alpha) uw.\label{bdef}
\end{equation}
Trajectories with four tangents $m^{\alpha}=(m^{u},m^{w},0,0)$
where
\begin{equation}
m^{u}=-2\,{\frac
{{\alpha}^{3}{M}^{2}\mathcal{B}}{\mathcal{A}{w}^{2}}}m^{w}\label{ucom}
\end{equation}
and
\begin{equation}
m^{w}=\delta{\mathcal{A}}^{4\,{\frac
{\alpha-3}{{\alpha}^{2}}}}{w}^{4\,{\frac {3
-\alpha}{{\alpha}^{2}}}}{e^{4\,{\frac {{\alpha}^{2}{M}^{2}(
3-\alpha)}{\mathcal{A}}}}},\label{wcom}
\end{equation}
where $\delta$ is a constant, are radial null geodesics. Now we
find that the expansion is given by
\begin{equation}
\nabla_{\alpha}m^{\alpha}=\frac{2\,\delta( \alpha-3 )
F{\mathcal{A}}^{-{\frac
{{\alpha}^{2}-4\,\alpha+12}{{\alpha}^{2}}}}{e^{4\,{\frac
{{\alpha}^{2}{M}^{2} ( 3-\alpha) }{\mathcal{A}}}} }\,}{{w}^{{
\frac { ( \alpha+6 )  ( \alpha-2 ) }{{\alpha}^{2 }}}}\mathcal{B}}
,\label{genexpansion}
\end{equation}
where
\begin{equation}
F= {\alpha}^{3}{M}^{2}( 2\,{\alpha}^{3}{M}^{2}-3\,wu ( \alpha- 2 )
) +uw(\mathcal{A}-{\alpha}^{4}{M}^{2}).\label{fdef}
\end{equation}
The choice $\alpha=3,\; \delta=1$ reproduces the Bertotti - Kasner
result and the choice $\alpha=2,\; \delta=\frac{16 M^2}{e}$
reproduces the Schwarzschild vacuum result both for $m^{\alpha}$
as given above. For $2 < \alpha < 3$ these geodesics can be given
explicitly in terms of elementary functions and are reproduced in
Appendix C. Some details are shown in FIG. \ref{schwarzdesitter}.

\begin{widetext}
\begin{figure}[ht]
\epsfig{file=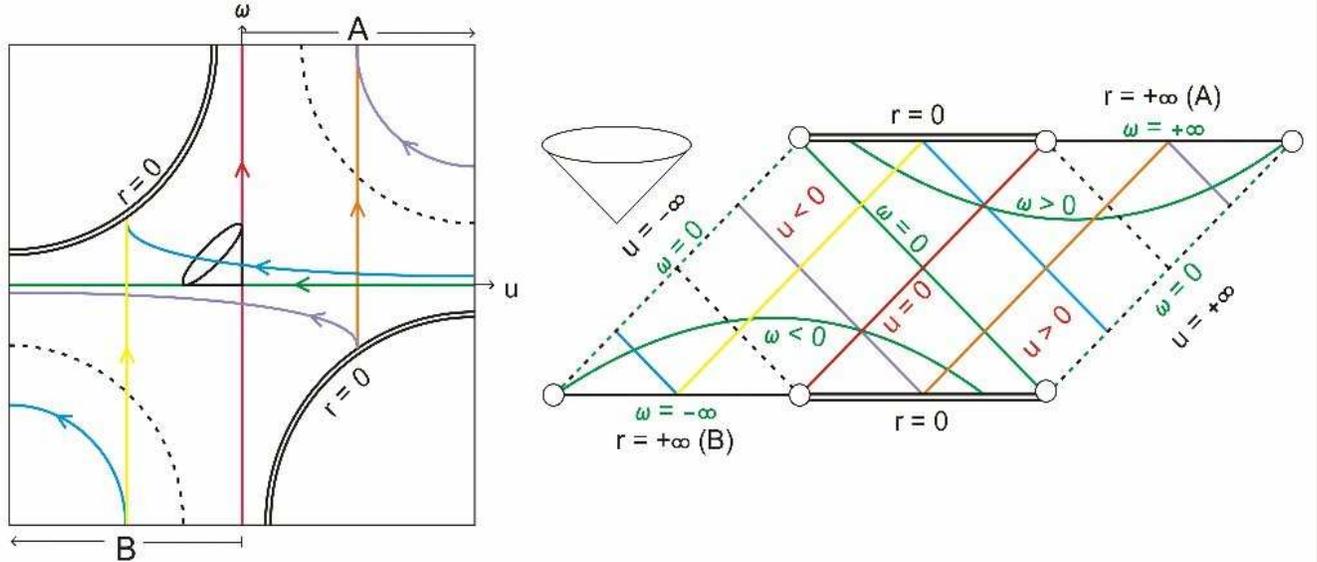,height=3in,width=7in,,angle=0}
\caption{\label{schwarzdesitter}(colour coded) At left,
representation of the Schwarzschild - de Sitter space in the $u -
w$ diagram. Coloured curves are radial null geodesics. The double
black curves give $r=0$  and the dashed black curves give the
cosmological horizons $r=E$. The axes are centered on $r=C$. The
future null cone at the origin ($u=w=0$) is shown. The
trajectories $w=$ constant are spacelike for finite $w\neq0$. At
right, the Penrose-Carter diagram of the Schwarzschild - de Sitter
space. The global orientation of the future null cone is shown.
The same scheme as the $u - w$ diagram is used and some
trajectories of constant $w$ are shown. Identification of $w=0,
u=\pm\infty$ gives a null geodesically complete causal
representation of the Schwarzschild - de Sitter space.}
\end{figure}
\end{widetext}

\subsection{$0< C < 2M$ (Schwarzschild anti - de Sitter)}
For $0 < \alpha < 2$ there are no cosmological horizons but
equations (\ref{adef}) through (\ref{fdef}) hold as given above.
Again the associated radial null geodesics can be given explicitly
in terms of elementary functions and are reproduced in Appendix D.
Some details are shown in FIG. \ref{antidesitterimage}.

\begin{widetext}
\begin{figure}[ht]
\epsfig{file=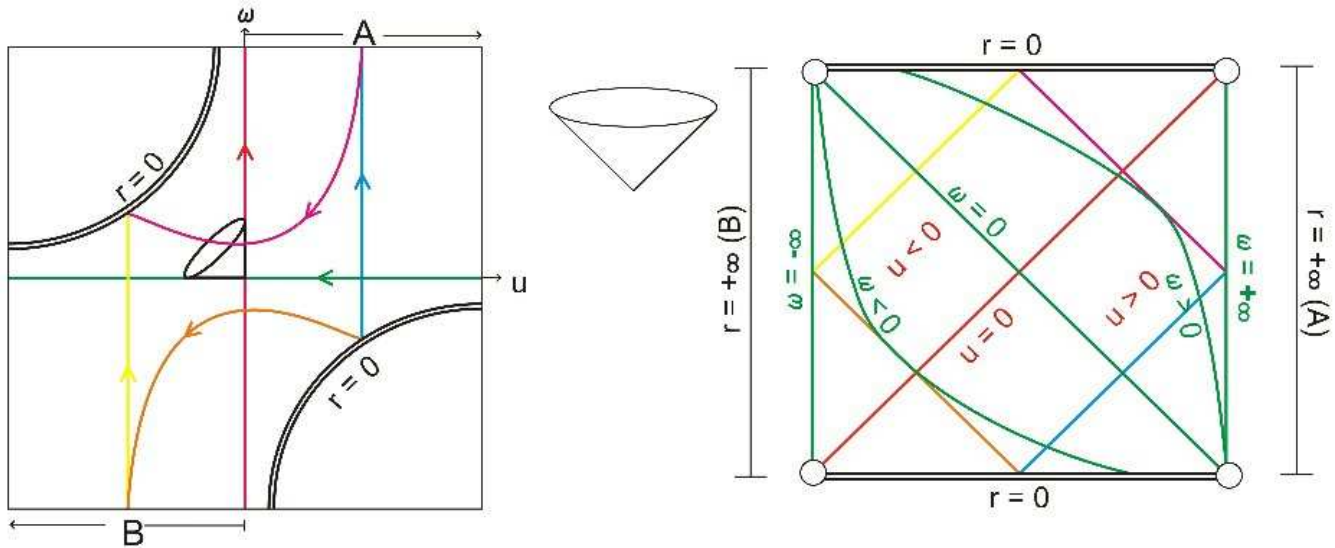,height=3in,width=7in,angle=0}
\caption{\label{antidesitterimage}(colour coded) At left,
representation of the Schwarzschild anti - de Sitter space in the
$u - w$ diagram. Coloured curves are radial null geodesics. The
double black curves give $r=0$.  The future null cone at the
origin ($u=w=0$) is shown. At right, the Penrose-Carter diagram of
the Schwarzschild anti - de Sitter space. The global orientation
of the future null cone is shown. The same scheme as the $u - w$
diagram is used.}
\end{figure}
\end{widetext}

\section{Coordinate construction (D $\geq$ 4)}

This section parallels II but extends the construction to
arbitrary dimensions ($D \geq 4$). Write
\begin{equation}
ds^2=fdu^2+2dudw+ r^2d\Omega_{D-2}^{2},\label{tangherlini}
\end{equation}
where $d\Omega_{D-2}^{2}$ is the metric of a unit $D-2$ sphere.
Again $k^{\alpha}=\delta^{\alpha}_{w}$ is tangent to null
geodesics affinely parameterized by $w$ but now the associated
expansion is given by
\begin{equation}
\nabla_{\alpha}k^{\alpha}=\frac{D-2}{r}\frac{ \partial r}{\partial
w}. \label{expansiongendg4}
\end{equation}
Further, associated with the vector field
$l^{\alpha}=\delta^{\alpha}_{u}$ we again have
$l^{\alpha}l_{\alpha}=f$ and
\begin{equation}
l^{\beta}\nabla_{\beta}l^{u}=-\frac{1}{2}\frac{
\partial f}{\partial w},\;\;\;\;l^{\beta}\nabla_{\beta}l^{w}=\frac{1}{2}(\frac{
\partial f}{\partial u}+f\frac{
\partial f}{\partial w}),\label{laccdg4}
\end{equation}
but now the associated expansion is
\begin{equation}
\nabla_{\alpha}l^{\alpha}=\frac{D-2}{r}\frac{
\partial r}{\partial u}.\label{lexpdg4}
\end{equation}
The trajectories for which all coordinates are constant except $u$
are radial null geodesics (affinely parameterized by $u$) only for
$f=\frac{
\partial f}{\partial u}=\frac{
\partial f}{\partial w}=0$.
Finally, as before, if $\frac{\partial f}{\partial u}=0$ then
$m^{\alpha}=(-2/f,1,0,0,...)$ is tangent to a radial null geodesic
with $w$ again affine and the associated geodesics immediately
follow as $u=-2 \int dw/f+ \delta$ where $\delta$ is a constant.

\section{Hyper-spherical vacua with $\Lambda$ (D $\geq$ 4)}

This section generalizes section III above. For dimensions $D \geq
4$ consider the spaces described by
\begin{equation}
ds^2=\frac{w^2 \psi_{D}(u,w)}{(r-C)^2(D-1)r^{D-3}C^3}du^2+2dudw+
r^2d\Omega_{D-2}^{2}, \label{metricsgen}
\end{equation}
where the function $\psi_{D}$ is given by
\begin{equation}
3(D-3)(C-2M)r^{D-3}(r-C)^{2}+2C( 3(D-3)M-(D-4)C)( {r}^{D-2}(D-3)
-{r}^{D-3}C(D-2)+C^{D-2})
 \label{fgen}
\end{equation}
and $r$ signifies the function
\begin{equation}
r \equiv \frac{( D-3) uw(3M-C)}{C^2}+C, \label{rfngen}
\end{equation}
with $C$ ($ \neq 0$) and $M$ ($ \geq 0$) constants. It follows
that \cite{Lambda}
\begin{equation}
R_{\alpha}^{\beta}=\Lambda
\delta_{\alpha}^{\beta},\label{riccigen}
\end{equation}
\begin{equation}
R=D\Lambda,\label{ricciscalargen}
\end{equation}
and
\begin{equation}
C_{\alpha \beta \gamma \delta}C^{\alpha \beta \gamma \delta}
=\frac{4(D-3)(D-2)^2C^{2(D-4)}((D-4)C-3(D-3)M)^2}{(D-1)r^{2(D-1)}},\label{weylscalargen}
\end{equation}
where
\begin{equation}
\Lambda=\frac{3(D-3)(C-2M)}{C^3}.\label{lambdagen}
\end{equation}
Again, the one independent invariant derivable from the Riemmann
tensor without differentiation can be taken to be
(\ref{weylscalargen}) and so with $u$ and $w$ extending over the
reals,  the spaces described by (\ref{metricsgen}) are maximally
extended and regular for $0< r < \infty$ (and for all $r$ if
$C_{\alpha \beta \gamma \delta}C^{\alpha \beta \gamma \delta}=0$,
see below). In particular, note that
\begin{equation}
g_{_{uu}}|_{_{r=C}}=\frac{w^2(D-3)((3M-C)D+5C-12M)}{C^3}
\label{regulargen}.
\end{equation}
Again viewing $\Lambda$ as a constant of nature and $M$ a property
of the vacuum, the specification of $C$ does not determine the
physical situation uniquely. This is discussed below. In view of
the generality of the cases considered, the formulae given are
remarkably simple.

\subsection{Comparison with curvature coordinates}
Consider the hyper-spherically symmetric spacetime
\begin{equation}
ds^2=-f(r)dt^2+\frac{dr^2}{f(r)}+r^2d\Omega_{D-2}^{2}.\label{curvature}
\end{equation}
Note that $r$ is now a coordinate. It is not difficult to show
that the unique form of $f$ which satisfies (\ref{riccigen}) is
given by
\begin{equation}
f=1-\frac{2\tilde{m}}{r^{D-3}}-\frac{\Lambda r^2}{D-1}
\label{curvaturef}
\end{equation}
where $\tilde{m}$ is a constant (which we take $\geq0$). This
solution was apparently first discussed by Tangherlini
\cite{tangherlinip}. The coordinates are, of course, defective at
$f=0$, but the associated roots depend on $D$ and are not so easy
to find from (\ref{curvaturef}). The previous construction
circumvents this unnecessary algebra as follows: For
(\ref{curvature}) with (\ref{curvaturef}) we find
\begin{equation}
C_{\alpha \beta \gamma \delta}C^{\alpha \beta \gamma \delta}
=\frac{4(D-2)^2((D-2)^2-1)\tilde{m}^2}{r^{2(D-1)}}
\label{weylsqcurv}
\end{equation}
and so from (\ref{weylscalargen}) and (\ref{weylsqcurv})
(assuming, without loss in generality, the same coordinates for
$d\Omega_{D-2}^{2}$)we have the relation
\begin{equation}
\tilde{m}^2=\frac{(D-3)C^{2(D-4)}((D-4)C-3(D-3)M)^2}{(D-1)((D-2)^2-1)}
\label{mrelation}
\end{equation}
which gives $\tilde{m}^2=M^2$ only for $D=4$. Now substituting for
$\tilde{m}$ from (\ref{mrelation}) into (\ref{curvaturef}) and
writing an associated root to $f=0$ as $r=C$ we obtain
(\ref{lambdagen}). In this way we never need to deal with
(\ref{curvaturef}) and so all dimensions $D \geq 4$ are equally
difficult to deal with as regards the location of horizons.

\subsection{Geometrical Mass}
The previous section raises the issue of ``mass". In spacetime
with $\Lambda=0$ the concept of mass in the spherically symmetric
case is well established \cite{mass}. The geometrical mass in $n$
dimensions for spaces admitting a $D-$sphere has been defined
previously by way of the sectional curvature \cite{lake1}. Whereas
this is merely a formal definition, it is of interest to compare
this definition with $M$ and $\tilde{m}$. In dimension $D$ define
\begin{equation}
\mathcal{M}\equiv\frac{1}{2}\;\;^{(D)}R_{\theta
\phi}^{\;\;\;\;\theta \phi}g_{\theta \theta}^{(D-1)/2}.
\label{geomass}
\end{equation}
In the special case $\Lambda=0 \;(C=2M)$ it follows that
\begin{equation}
\mathcal{M}=\tilde{m}=2^{D-4}M^{D-3}. \label{geomassnolambda}
\end{equation}
More generally, for $\Lambda\neq0 \;(C\neq2M)$, it follows that
\begin{equation}
\mathcal{M}=\tilde{m}+\frac{\Lambda r^{D-1}}{2(D-1)},
\label{geomasslambda}
\end{equation}
where $\Lambda$ is given by (\ref{lambdagen}).  We can, of course,
replace $\tilde{m}$ by $M$ in (\ref{geomasslambda}) by solving for
$M$ from (\ref{mrelation}). We now consider some specific cases.

\subsection{de Sitter space  (D $\geq$ 4)}
We define hyper-spherical de Sitter space by the requirement
$C_{\alpha \beta \gamma \delta}C^{\alpha \beta \gamma \delta}=0$
so that from (\ref{weylscalargen}) and (\ref{weylsqcurv}) we have
\begin{equation}
M=\frac{C(D-4)}{3(D-3)},\;\;\;\;\tilde{m}=0 \label{desittergen}
\end{equation}
and so
\begin{equation}
\Lambda=\frac{D-1}{C^2}. \label{lambdadesitter}
\end{equation}
The metric follows as
\begin{equation}
ds^2=\frac{\Lambda}{D-1}w^2du^2+2dudw+\frac{\Lambda}{D-1}(\frac{D-1}{\Lambda}-uw)^2d\Omega_{D-2}^{2}.
\label{desittergenmetric}
\end{equation}
Trajectories with four tangents
$m^{\alpha}=(-\frac{2(D-1)}{\Lambda w^2} ,1,0,0,...)$ are radial
null geodesics (so $w$ is affine for both radial null directions).
We can write these geodesics in the form
\begin{equation}
w=\frac{2(D-1)}{\Lambda u+\delta}\label{desittergeogen}
\end{equation}
where $\delta$ is a constant. The negative cosmological horizons
($r=-\sqrt{\frac{(D-1)}{\Lambda}}$) are then given by $\delta=0$
where the expansion $\nabla_{\alpha}m^{\alpha}$
($=\frac{(D-2)(2(D-1)-uw \Lambda )}{w((D-1)-uw \Lambda)}$)
vanishes. The associated diagram is qualitatively the same as FIG.
\ref{desitterfig}.

\subsection{$C=3M$ (Bertotti-Kasner space) (D $\geq$ 4)}

With $C=3M=\sqrt{\frac{D-3}{\Lambda}}=r$ we have hyper-spherical
Bertotti - Kasner space. The metric simplifies to
\begin{equation}
ds^2=\Lambda w^2
du^2+2dudw+\frac{D-3}{\Lambda}d\Omega_{D-2}^{2}.\label{bertottigen}
\end{equation}
Trajectories with four tangents $m^{\alpha}=(-\frac{2}{\Lambda
w^2},1,0,0,...)$ are radial null geodesics (so again $w$ is affine
for both radial null directions) and again
$\nabla_{\alpha}k^{\alpha}=\nabla_{\alpha}m^{\alpha}=0$. We can
also write these geodesics in the form (\ref{bertottigeo}).

\subsection{Tangherlini black holes (D $\geq$ 4)}
Asymptotically flat static vacuum black holes (admitting the $D-2$
sphere) are unique \cite{flat} and given by the Tangherlini
generalization of the Schwarzschild vacuum \cite{tangherlinip}. In
our notation these correspond to the case $C=2M$. Global, regular
and explicit coordinates for these spaces have been given
previously \cite{lake}. Some further discussion is given in
Appendix E. More generally, the radial null geodesics that satisfy
\begin{equation}
\frac{d w}{d u}=-\frac{1}{2}\frac{w^2
\psi_{D}(u,w)}{(r-C)^2(D-1)r^{D-3}C^3},\label{othernullhigh}
\end{equation}
with $\psi_{D}$ given by (\ref{fgen}) and $r$ by (\ref{rfngen}),
are of interest. For these trajectories, excluding the cases
discussed above, it is clear that $du/dw \rightarrow 0$ as $r
\rightarrow 0$ and $dw/du \rightarrow 0$ as $w \rightarrow 0$
($u\neq0$). From (\ref{lambdagen}), writing $C=\alpha M$ we have
$\Lambda M=3(D-3)(\alpha-2)/\alpha^3$. In the range $2 < \alpha <
3$ there is another constant $\beta > \alpha$ for which $\Lambda
M=3(D-3)(\beta-2)/\beta^3$. The trajectories $r=\beta M$ are
radial null geodesics and are tangent to the ``cosmological"
horizons. It is also clear that for $C < 2M$ the solutions to
(\ref{othernullhigh}) evolve in a fundamentally different way as
in the case $D=4$. In fact, the qualitative behavior of all
solutions to (\ref{othernullhigh}) can be obtained without
explicit integration. The essential conclusion is that the Figures
given for $D=4$ hold, qualitatively, for $D>4$. This is discussed
in detail elsewhere \cite{details}.

\section{Summary}
A maximally extended, explicit and regular covering of the
Schwarzschild - de Sitter vacua in arbitrary dimension ($D \geq
4$) has been given. It has bee stressed that these coordinates
offer important advantages over the Kruskal - Szekeres procedure
that include: An explicit representation of the line element that
can be extended to arbitrary dimension, a simultaneous covering of
both the black hole and cosmological horizons and derivation by
direct integration of Einstein's equations without recourse to
coordinate transformations. In view of the generality of the
problem solved, the resultant formulae obtained are of a
remarkably simple form.
\begin{acknowledgments}
Part of this work was reported at the ``The Dark Side of Extra
Dimensions" workshop held in May 2005 at the Banff International
Research Station (BIRS), Banff, Alberta. It is a pleasure to thank
Werner Israel and Don Page for comments and Taylor Binnington for
drawing the figures. This work was supported by a grant from the
Natural Sciences and Engineering Research Council of Canada and
was made possible by use of \textit{GRTensorIII} \cite{grt}.
\end{acknowledgments}

\begin{appendix}

\section{anti - de Sitter space (D = 4)}

Anti - de Sitter space is not contained in (\ref{metric}) with
(\ref{r(u,w)}) but can be given in the form (\ref{generalmetric})
with $h=1$ and is included here for completeness. It is given by
\begin{equation}
ds^2=(\frac{w^2
\Lambda}{3}-1)du^2+2dudw+w^2d\Omega^2.\label{antidesitter}
\end{equation}
where the constant $\Lambda<0$. Equations (\ref{ricci}) and
(\ref{ricciscalar}) of course hold and the space is conformally
flat. Trajectories with four tangents
$m^{\alpha}=(\frac{6}{\Lambda w^2-3},-1,0,0)$ are radial null
geodesics so $w$ is again affine. We can write these geodesics in
the form
\begin{equation}
w=\sqrt{\frac{3}{-\Lambda}}\tan(\frac{1}{2}\sqrt{\frac{-\Lambda}{3}}(u+\delta))\label{antidesittergeo}
\end{equation}
where $\delta$ is a constant. With $u$ and $w$ extending over the
reals,  (\ref{antidesitter}) is maximally extended and regular.

\section{Degenerate Schwarzschild - de Sitter space (D = 4)}

\begin{widetext}
\begin{figure}[ht]
\epsfig{file=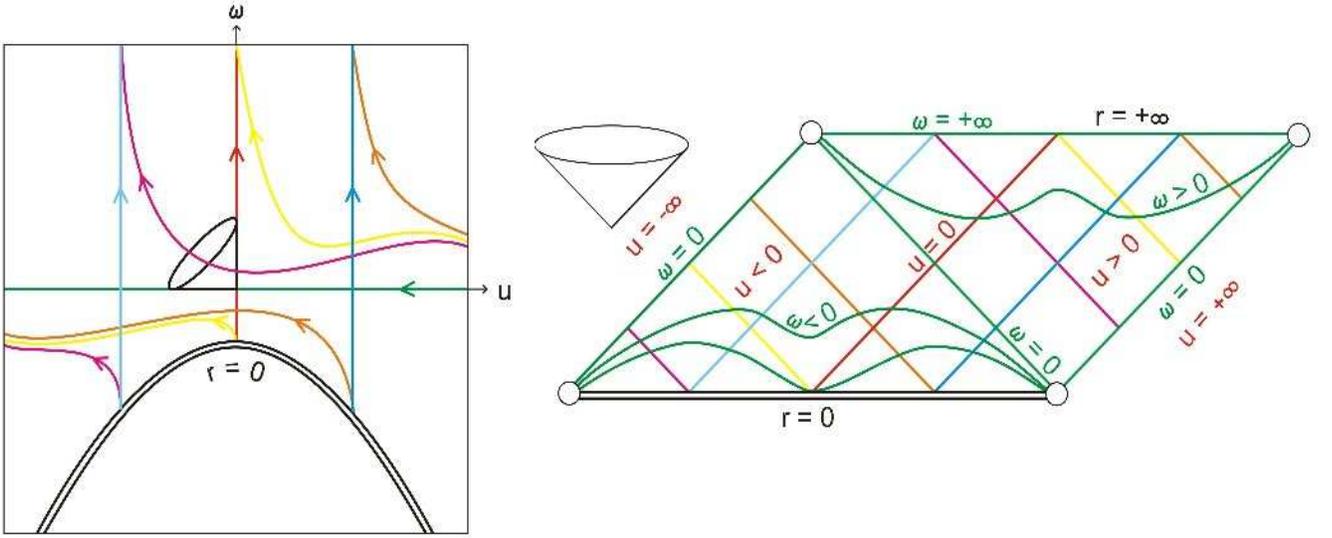,height=3in,width=7in,angle=0}
\caption{\label{degenschwarzdesitterimage}(colour coded) At left,
representation of the degenerate Schwarzschild - de Sitter space
in the $u - w$ diagram. Coloured curves are radial null geodesics.
The double black curve gives $r=0$. The degenerate horizon occurs
at $w=0$. The future null cone at the origin ($u=w=0$) is shown.
At right, the Penrose-Carter diagram of the degenerate
Schwarzschild - de Sitter space. The global orientation of the
future null cone is shown. The same scheme as the $u - w$ diagram
is used and some trajectories of constant $w$ are shown. Shown is
the ``white hole" case. The black hole case is obtained by a flip
in the vertical so null geodesics terminate (as opposed to
originate) at $r=0$. }
\end{figure}
\end{widetext}

The degenerate Schwarzschild - de Sitter black hole has
$3M=\frac{1}{\sqrt{\Lambda}}$ like the Bertotti-Kasner space but
it is not contained in (\ref{metric}) with (\ref{r(u,w)}) but can
be given in the form (\ref{generalmetric}) with $h=1$ and is
included for completeness. Coordinates for degenerate black holes
are seldom discussed \cite{degen}. Now
\begin{equation}
f={\frac {-w ( 12\,u r^{2}
 ( {u}^{2}+1 ) +w ( w+3\,{u}^{2}+3 )  ( w+{
u}^{2}+1 )  ) }{ 3r ^{2}
 ( {u}^{2}+1 ) ^{2}}}
\label{degenf}
\end{equation}
with
\begin{equation}
r={\frac {w+{u}^{2}+1}{\sqrt {\Lambda} ( {u}^ {2}+1 ) }}
.\label{degenr}
\end{equation}
In addition to the radial null geodesics given by constant $u$ and
$w=0$, the other radial null geodesics can be given by
\begin{equation}
w=\frac{3(u^2+1)}{2\mathcal{W}(k,\pm x)-1}\label{degennull}
\end{equation}
where $\mathcal{W}$ is the Lambert W function \cite{lambert} with
$k=(0(+),-1(-))$ and
\begin{equation}
x=\frac{1}{4} \Lambda u(u^2+3)+\delta \label{degennullpsi}
\end{equation}
where $\delta$ is a constant. Some details are shown in FIG.
\ref{degenschwarzdesitterimage}.

\section{Schwarzschild - de Sitter geodesics (D = 4)}
For $2 < \alpha < 3$ the radial null geodesics with tangents
$m^{\alpha}$ can be given in a simple form: $GH$ is a constant of
motion where
\begin{equation}
G=\left( { \frac {{M}^{2}{\alpha}^{2} ( 2\,\alpha+\beta ) +uw( 3
-\alpha ) }{uw( \alpha-3 ) +{M}^{2}{\alpha}^{2} ( -\alpha+\beta )
}} \right) ^{( 6-\alpha ) \alpha} \label{schwdsgeo}
\end{equation}
and
\begin{widetext}
\begin{eqnarray}
H=( {M}^{2}( {M}^{2}{\alpha}^{3}( 2\,{\alpha}^{3}{M}^{ 2}-3\,uw (
-2+\alpha )  ) +{u}^{2}{w}^{2}( -2+ \alpha ) ( \alpha-3 ) +{w}^{2}
) ) ^{ ( 2-\alpha ) ( \alpha+2\,\beta ) }. \label{Gdef}
\end{eqnarray}
\end{widetext}

\section{Schwarzschild - anti de Sitter geodesics (D = 4)}
Writing $C=M/\gamma$ with $\gamma>1/2$ the radial null geodesics
with tangents $m^{\alpha}$ can be given in the form
\begin{widetext}
\begin{eqnarray}
2\,\arctan ( {\frac { ( 2\,\gamma-1 )  ( 2\,u{ \gamma}^{2} (
3\,\gamma-1 ) w+3\,{M}^{2} ) }{\sqrt {
 ( 6\,\gamma+1 )  ( 2\,\gamma-1 ) }{M}^{2}}}
 )  ( 6\,\gamma-1 )\\ =\ln  ( {\frac {{M}^{2}
 ( 2\,{M}^{2}+3\,{\gamma}^{2}uw ( 2\,\gamma-1 )
 ) +{u}^{2}{\gamma}^{4}{w}^{2} ( 3\,\gamma-1 )
 ( 2\,\gamma-1 ) }{w}} ) \sqrt { ( 6\,\gamma+1
 )  ( 2\,\gamma-1 ) }+\delta
 \label{antigeo}
\end{eqnarray}
\end{widetext}
where $\delta$ is a constant.

\section{Schwarzschild-Tangherlini black holes (D $\geq$ 4)}
In the present notation we set $C=2M$ and so have
(\ref{tangherlini}) with
\begin{equation}
f={\frac {w^{2} ((D-3)r+2M(2-D)+(2M)^{D-2}r^{3-D})}{2M
(r-2M)^{2}}},\label{tangherlinif}
\end{equation}
where
\begin{equation}
r= \frac{(D-3)uw}{4M}+2M. \label{rfn}
\end{equation}
With $D=4$ we recover (\ref{schw}).  Even implicit forms of $r$ in
the Kruskal - Szekeres procedure are not known for for all $D$
(\textit{e.g} $D=8,10$ \cite{lake}). However here we need only
substitute for $D$ and insert (\ref{rfn}) into
(\ref{tangherlinif}) to give a maximally extended, explicit and
regular coverings of the space.  For example, with $D=5$ we have
\begin{equation}
f={\frac {2{w}^{2} ( 6\,{M}^{2}+uw ) }{ ( 4\,{M}^{2}+u w ) ^{2}}}.
\end{equation}
Solutions to (\ref{othernull}) with (\ref{tangherlinif}) and
(\ref{rfn}) can be given explicitly in terms of elementary
functions in some cases. For example, with $D=5$ we have

\begin{equation}
uw=-2M^2(\mathcal{W}(\delta w)+4)
\end{equation}
where $\delta$ is a constant.
\end{appendix}


\begin{thebibliography}{}
\bibitem{transformations}Israel \cite{israel} (in the published form of his work) introduced the coordinates by way
of a transformation which, in his notation reads as $t=r-2m
\ln(u/(8m^2z))$. In this paper Einstein's equations for vacua have
been solved directly without prior coordinates. This is not
possible for Kruskal - Szekeres coordinates wherein even at $D=4$
with $\Lambda = 0$ the vacuum field equations are difficult to
solve. See, for example, Lake \cite{procedure}.
\bibitem{conventions}We use Einstein's equations (with cosmological constant),
geometrical units and a signature of $D-2$. The sign conventions
for $w$ and $u$ is that $w$ can decrease and $u$ increase only in
spacelike directions. For convenience, explicit functional
dependence is usually shown only on the first appearance of a
function.
\bibitem{Lambda} If the cosmological constant $\lambda$ is defined (in
dimension $D$) by the vacuum field equations $^{(D)}G_{\alpha
\beta}+\lambda g_{\alpha \beta}=0$ where $^{(D)}G_{\alpha \beta}$
is the $D$ dimensional Einstein tensor, then $\Lambda=2
\lambda/(D-2)$.
\end{thebibliography}
\end{document}